\documentclass[twocolumn,twoside,slac_two]{revtex4}
\usepackage{graphicx}
\usepackage{fancyhdr}
\newcommand\pflux{ph cm$^{-2}$ s$^{-1}$}

\begin{document}

\title{General Properties of \textit{Fermi}/LAT Active Galactic Nuclei}

%

\author{B. Lott} 
\affiliation{Univ. Bordeaux, CENBG, UMR 5797, F-33170 Gradignan, France}
\affiliation{CNRS, IN2P3, CENBG, UMR 5797, F-33170 Gradignan, France}
\author{E.~Cavazzuti, S.~Cutini,  D.~Gasparrini}
\affiliation{Agenzia Spaziale Italiana (ASI) Science Data Center, I-00044 Frascati (Roma), Italy}
\author{C.~D.~Dermer}
\affiliation{Space Science Division, Naval Research Laboratory, Washington, DC 20375-5352}
\author{on behalf of the \textit{Fermi}/LAT Collaboration}

\begin{abstract}
The Second Catalog of Blazars and other Active Galactic Nuclei (AGNs) detected by the \textit{Fermi}/LAT (2LAC) includes about 1100 sources, 886 of which comprise the Clean Sample.  The general properties of the different populations of sources classified according to the strength of their emission lines (FSRQs, BL~Lacs) or the estimated position of the synchrotron peak are reviewed. \end{abstract}

\maketitle

\thispagestyle{fancy}
\section{Introduction}

Since the launch of the \textit{Fermi} satellite in June 2008, the \textit{Fermi}-LAT has opened a new era in high-energy astrophysics. The LAT instrument was built by an international consortium of institutions from the USA, Italy, France, Sweden and Japan. It is composed of 16 elements each including a tracker, enabling the direction of the incident photon to be reconstructed, and a calorimeter, measuring the energy of the electromagnetic shower initiated by the gamma-ray photon. The whole instrument is surrounded by an anticoincidence detector allowing a discrimination between gamma-ray photons and background charged cosmic rays. The LAT energy range is 20 MeV--300 GeV, the high-energy end corresponding to a statistical shortage of detected photons. The point-spread function of the detector is strongly energy dependent. It is governed by the multiple scattering of the initial electron-positron pair at low energy and by the pitch of the silicon strips at high energy. The field of view is 2.4 sr, i.e., about 20\% of the entire sky is covered at a given time. The orbit altitude is 565 km, corresponding to a period of about 91 minutes. In survey mode, the instrument rocks by about 50$^\circ$ altenatively south and north over one orbit, enabling the whole sky to be scanned every 3 hours with roughly uniform exposure.  The mission has a guaranted 5-year lifetime, extending till 2013. At the end of that period, a 5-year source catalog will be produced. The instrument was designed with a target livetime of 10 years,  with no consumable aboard. So far the deterioration rate of the detector has been extremely slow, with fewer than 500  (out of 880 000) tracker strips disabled. Every 2 years (starting in 2012) a review panel will issue recommendations about extending the support to the mission, the first examined period being 2013--2014.

\section{The variable sky seen by the LAT}
\begin{figure*}
   \centering
   \includegraphics[width=12cm]{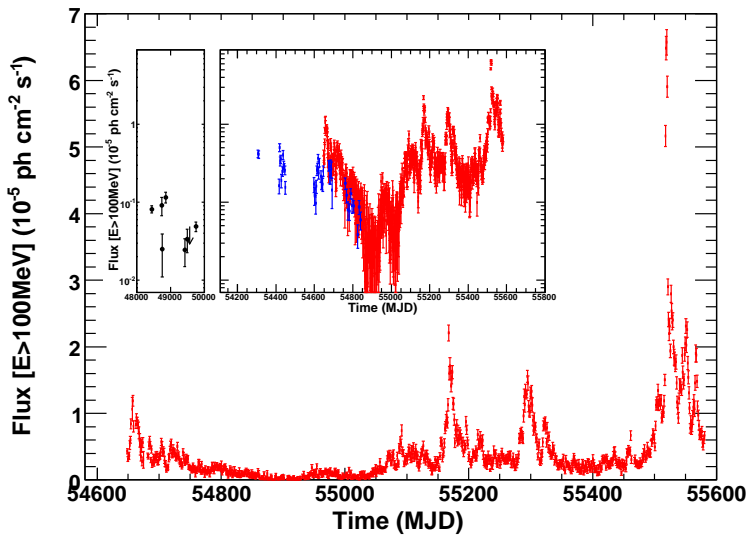}
   \caption{ Daily light curve of 3C454.3 measured with the Fermi-LAT since launch. Inset: Historical light curve. Black points are from EGRET \cite{Har99} and blue points are from AGILE \cite{Str10}.
            \label{fig:3C454}
           }
    \end{figure*}

The LAT presents many assets regarding AGN Science. Its unprecedented sensitivity, almost uniform at high Galactic latitude, makes it very suitable to study populations of extragalactic objects. The continuous scan in survey mode allows alerts to be issued shortly after transient or new flaring sources are detected and enables as well source monitoring on time scales ranging from months down  to a few hours. One of the most spectacular examples concerns the flares of the blazar 3C~454.3 (Fig. \ref{fig:3C454}), which reached a record flux level above 100 MeV of 8 $\times 10^{-5}$ \pflux, ($\gamma$-ray luminosity $>$ 10$^{50}$ erg/s) showing variability time scales of a few hours \cite{3C454_11}. 
 
Another breakthrough arises from the \textit{Fermi}/LAT large energy range extending into the 10-100 GeV domain, which was previously little explored. This capability has allowed new spectral features at high energy to be discovered  and has also enabled the identification of potential candidates of TeV sources, effectively leading to several discoveries by ground-based Cherenkov detectors. 

At the time of this conference, about 130 Astronomers Telegrams (ATels) have been issued on the basis of LAT data, of which 120 concerned AGNs. The alert threshold is a flux above 100 MeV greater than $10^{-6}$ \pflux, but other considerations like a significant flux rise with respect to the average flux, the detectability at TeV energies for a known source or the rarity of a transient event come into play when issuing an Atel. The ``Flare Advocates'' constantly monitor the sky, issue alerts and feed the Fermi blog. The list of ATels is posted at \url{http://www-glast.stanford.edu/cgi-bin/pub_rapid}.         

As EGRET demonstrated that the extragalactic $\gamma$-ray sky was dominated by blazars, there were great expectations concerning advances in blazar and AGN science from the \textit{Fermi}/LAT. Many of these expectations have been or are about to be fulfilled. In the following, recent advances concerning \textit{Fermi}/LAT blazar populations and their different properties are presented. 

\section{Blazar Populations}

The Second Catalog of Active Galactic Nuclei Detected by the \textit{Fermi}/LAT \cite{2LAC}, referred to as 2LAC in the following is a spinoff of the \textit{Fermi}/LAT Second Source Catalog (2FGL) \cite{Cat} based on data accumulated over the first 24 months of sky survey.  A first list of \textit{Fermi}/LAT-detected AGNs using  3 months of data and including 106 high-confidence sources, called the LAT Bright AGN Sources (LBAS) \cite{LBAS} was released in early 2009. The First LAT AGN Catalog (1LAC) \cite{1LAC} was released in the Fall of 2010. It  made use of 11 months of data and included 709 sources.  

The mean 95\% containment radius is 0.15$^\circ$ for the high-latitude 2FGL sources. The association of a given source with an AGN is based on spatial coincidence following three different approaches described in \cite{2LAC}.  Different catalogs were used for this purpose including  the Combined Radio All-sky Targeted Eight GHz Survey (CRATES) \cite{Hea07}, the Candidate Gamma-Ray Blazar Survey (CGraBs) \cite{Hea08} and the Roma-BZCat \cite{Mas09}.  This procedure leads to the association of 1017 sources at $|b|>10^\circ$ (with association probability $P_i>$0.8) corresponding to 991 distinct $\gamma$-ray sources (out of 1319 2FGL sources lying in this galactic-latitude range). A total of 886 sources belong to the ``Clean Sample'' meeting the following conditions: no double associations and no anomaly observed in the association or detection procedures.  Only roughly 12 associations are expected to be spurious in that sample. The validity of the procedure is confirmed by the good agreement of the actual distribution of angular separation between $\gamma$-ray sources and their AGN counterparts with theoretical expectations. 
\begin{figure*}
   \centering
   \includegraphics[width=15cm]{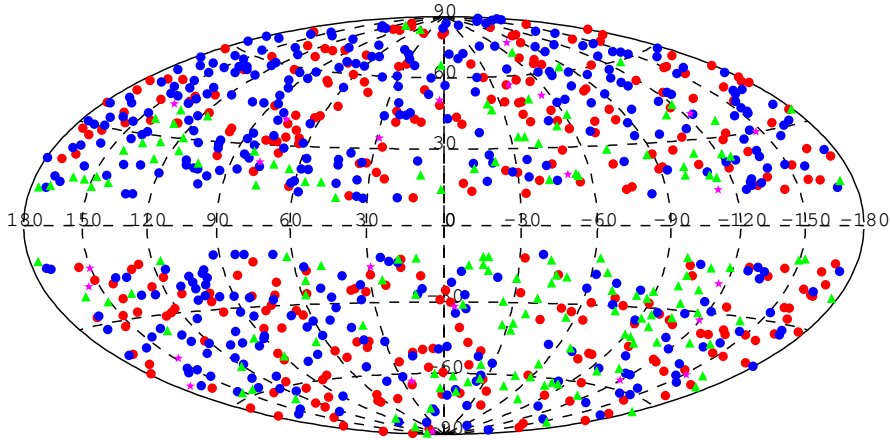}
   \caption{Locations of sources in the 2LAC Clean Sample.  Red:\ FSRQs, blue:\ BL~Lacs, magenta:\ radio galaxies, green:\ AGNs of unknown type.
            \label{fig:sky}
           }
    \end{figure*}

Although the fraction of identification of high-galactic sources is fairly large ($\simeq$75\%), some identifications are clearly missed as a significant asymmetry between northern and southern galactic hemispheres is observed (only 38\% of BL~Lacs are observed at $b<-10^\circ$,  while little asymmetry is observed for the whole 2FGL sources), manifesting the incompleteness of the parent catalogs.

The classification in terms of Flat-Spectrum Radio Quasars (FSRQs) and BL Lac-type objects (BL~Lacs) is conventionally based on the strength of the emission lines. This classification has been used in both the 1LAC and the 2LAC. In addition, a more refined  classification based on the source spectral energy distribution (SED) and  dealing with the location of the synchrotron peak $\nu_{syn}$ has been carried out by using archival data at lower wavelengths, more precisely $\alpha_{ro}$ and $\alpha_{ox}$, the radio-to-optical and optical-to-X-ray spectral slopes. The empirical relation between ($\alpha_{ro}$, $\alpha_{ox}$) and $\nu_{syn}$ was derived from a set of 48 high-quality, simultaneous SEDs of $\gamma$-ray blazars \cite{SED}. Sources were classified as low-, intermediate- or high-synchrotron peaked sources (referred to as LSPs, ISPs, HSPs respectively) if $\log$($\nu_{syn}/Hz$)$<$14, 14$<\log$($\nu_{syn}/Hz$)$<$15 or $\log$($\nu_{syn}/Hz$)$>$15 respectively. 

The census of sources in the 2LAC Clean Sample is as follows: 310 FSRQs, 395 BL~Lacs (44\% of which have a measured redshift), 24 other AGNs, 157 of unknown type. The more frequent detections of BL~Lacs relative to FSRQs is in sharp contrast with the situation encountered with EGRET \cite{Har99} where FSRQs outnumbered BL~Lacs by a factor $\simeq$ 3. This is probably primarily related to the larger sensitivity at high energy of the LAT, as most BL~Lacs  have harder spectra than FSRQs as discussed below. Moreover, the LAT is far more sensitive to multi-GeV photons than was EGRET, which lost sensitivity above several GeV due to self-vetoing effects.

The 24 non-blazar objects in the 2LAC Clean Sample include 8 misaligned AGNs, 4 Narrow-Line Seyfert 1 galaxies, 10 AGNs of other types and 2 starburst galaxies. A total of 45 (out of 599) 1 LAC Clean Sample sources are missing in 2LAC.
Due to variability,  misaligned AGNs as 3C 78, 3C 111 and 3C 120 don't make the significance cut over the 24 month period covered by the 2LAC while they were present in the 1LAC. Fornax A and Cen B are new entries relative to 1LAC.  

 All (except for a few) FSRQs for which $\nu_{syn}$ can be estimated from archival data are of the LSP type. For BL Lacs, the breakdown is 61 LSPs, 81 ISPs, and 160 HSPs. In line with the trend already observed in the BL~Lac/FSRQ ratio, HSPs constiture the most abundant subclass.

\begin{figure*}
\centering
\includegraphics[width=10cm]{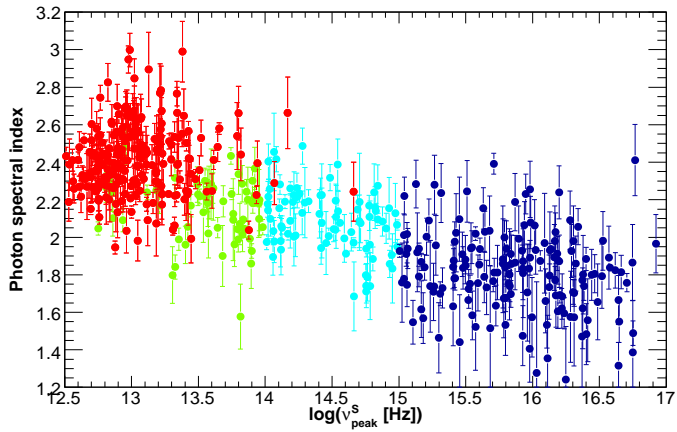}
\caption{Peak frequency of the SED synchrotron component vs. photon spectral index  for FSRQs (red) and BL~Lacs
(green:\ LSPs, cyan:\ ISPs, blue:\ HSPs) in the 2LAC Clean Sample.}
\label{fig:index_log_nu_syn}
\end{figure*}

The LAT-detected FSRQs all
exhibit soft \mbox{$\gamma$-ray} spectra while BL~Lacs are more diverse as illustrated in Figure~\ref{fig:index_log_nu_syn}, showing the spectral photon index measured with the LAT  plotted versus the estimated position of the synchrotron peak.

\subsection{Redshift distribution}     
\begin{figure*}
 \centering
\includegraphics[width=10cm]{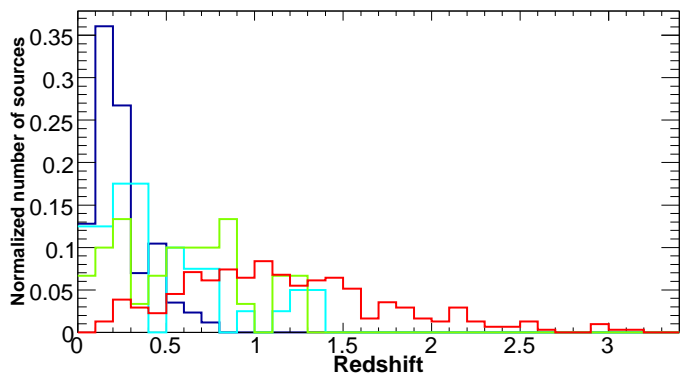}
\caption{Normalized redshift distribution for the FSRQs  (red),
LSP-BL~Lacs (green), ISP-BL~Lacs (cyan) and HSP-BL~Lacs (blue) in the Clean Sample.}
\label{fig:redshift}
\end{figure*}
          
The redshift distributions for FSRQs and BL~Lacs in the Clean Sample are
presented in Figure~\ref{fig:redshift}.  Note that only 44\% of 2LAC BL~Lacs
have measured redshifts.  This is notably worse than for LBAS, in which 29 out
of 42 (69\%) BL~Lacs had measured redshifts, but similar to 1LAC (42\%).   The highest redshift for a high-confidence 2LAC FSRQ is $z=3.10$.  For comparison, the maximum redshift in both CGRaBS and Roma-BZCAT is for an FSRQ with $z > 5$. No strong evidence for a new population of misaligned FSRQs emerging at lower redshifts is found.  

It is interesting to compare the redshift distribution of the
LAT blazars (in particular, the FSRQs) with that of the objects detected by BAT
on {\it Swift} \cite{ajello09}. In  the BAT survey, 40\% of all FSRQs have $z > 2$, and the distribution extends to
$z \sim 4$.  This behavior may be indicative of a shift in the SED peak
frequencies toward lower values (i.e., a ``redder'' SED) for blazars at high
redshifts.  Indeed, the jets of the high-redshift BAT blazars are found to be
more powerful than those of the LAT blazars and are among the most powerful
known \cite{Ghi10}.  The peak of the inverse Compton flux for these objects,
estimated to be in the MeV or even sub-MeV range, is located closer to the BAT
band than to the LAT band, and the LAT instead samples the cutoff region of the
inverse Compton spectrum.

\begin{figure*}
\begin{center}
\includegraphics[width=10cm]{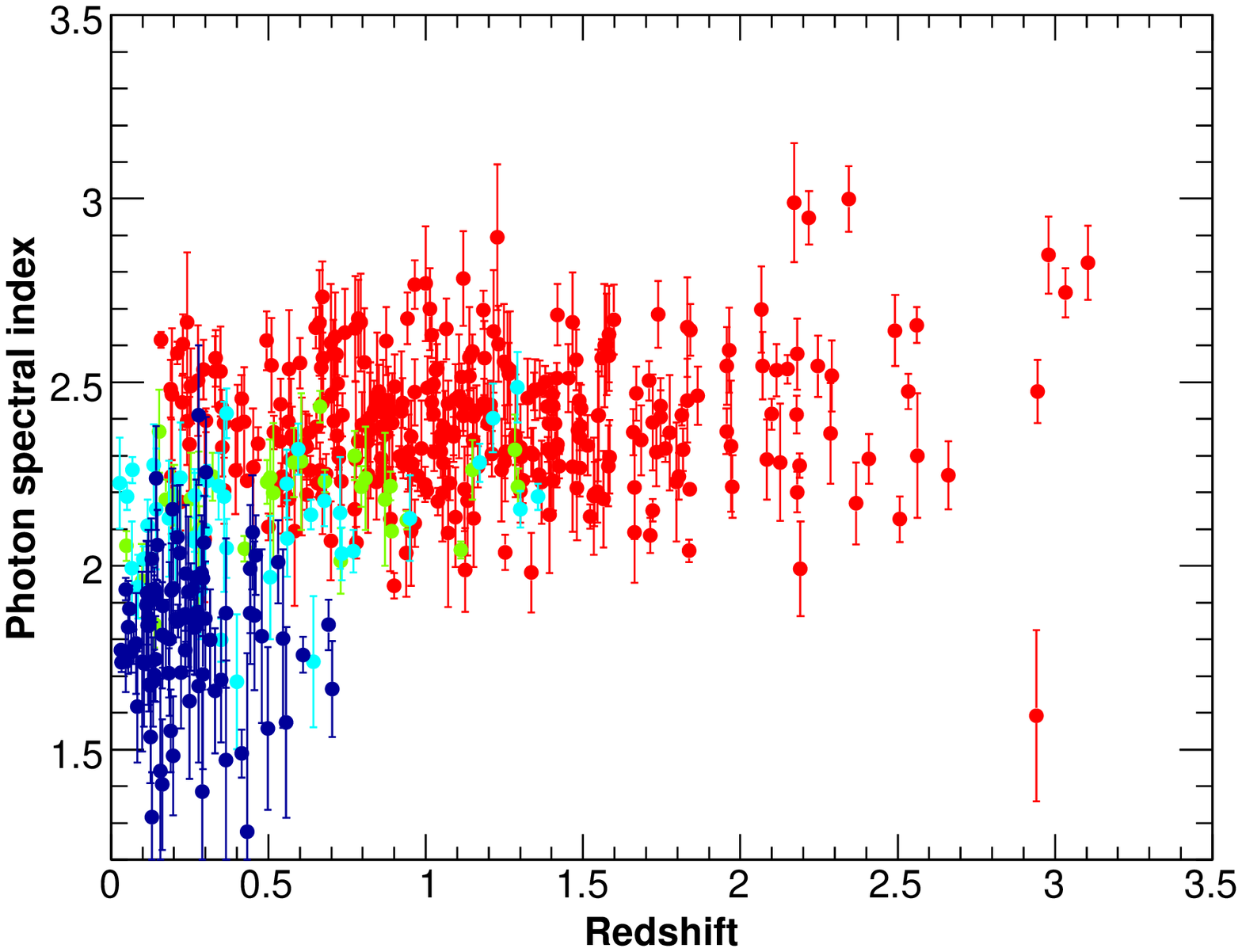}
\end{center}
\caption{Redshift vs. photon spectral index vs. for FSRQs (red) and BL~Lacs (green:\ LSPs, cyan:\ ISPs, blue:\ HSPs) in the Clean Sample.}
\label{fig:index_redshift}
\end{figure*}

The photon spectral index is plotted against redshift in
Figure~\ref{fig:index_redshift}.  For FSRQs, no significant evolution is
visible.  This behavior is compatible with what was previously observed for
LBAS and 1LAC.  The attenuation effect of the extragalactic background light (EBL) would
tend to introduce spurious evidence of evolution, but the soft
spectra of FSRQs and the common presence of spectral breaks at a few GeV
\cite{Spec} both minimize this effect.  A stronger evolution is seen for
BL~Lacs:\ hard sources are mostly located at low redshifts, while most
high-redshift sources are softer than average (though it is important to bear
in mind that most BL~Lacs do not have measured redshifts).  

\subsection{Spectral break}
While all AGNs spectra measured with EGRET were consistent with single power law distributions, the extended energy range covered by the \textit{Fermi}/LAT has enabled to discover the presence of a spectral break at a few GeV for bright LBAS FSRQs and some  LSP-/ISP-BL~Lacs \cite{Spec}. So far, no break has ever been found in the spectrum of any bright HSP-BL~Lacs, while if present it would have easily be detectable given the relative hardness of these sources. The first evidence for such feature was found in 3C 454.3 \cite{3C454}, which is still so far the best studied case thanks to its high recent brightness. A good model to the data is a broken power law with a break energy of 2.5 GeV and a change in spectral slope of about 1. The break origin is still unknown, several possibilities  are being actively discussed: feature if the underlying electron energy distribution \cite{3C454}, superposition of two external Compton components \cite{Fin10}, attenuation due to pair production with He II Ly-continuum photons from the broad-line region \cite{Pou10}. 
 
In the 2LAC Clean Sample, 69 bright sources (57 FSRQs and 12 BL~Lac LSPs) show very significant curvature.  One must keep in mind that in order to show a high curvature index a source needs both to have a spectrum deviating from a power-law distribution and be bright enough so that the observed deviation is statistically significant. Some fainter sources thus probably do exhibit curved spectra as well.

\section {Luminosity Distributions}

\begin{figure}
\centering
\includegraphics[width=9cm]{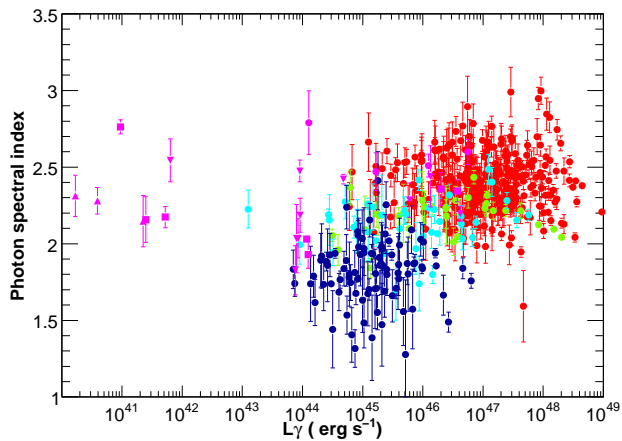}
\caption{\mbox{$\gamma$-ray} luminosity vs. photon spectral index  for the different AGN classes (red:\ FSRQs, blue:\ BL~Lacs, magenta:\ non-blazar AGNs, green:\ AGNs of unknown type) in the Clean Sample.}
\label{fig:index_lum}
\end{figure}

Figure~\ref{fig:index_lum} shows the photon spectral index plotted against the
\mbox{$\gamma$-ray} luminosity. This correlation has been widely discussed in the context of the ``blazar divide'' \cite{Ghisellini09}. It is important to bear in mind important caveats when interpreting this correlation.  For instance, more than half of the BL~Lacs lack measured redshifts.  Strongly beamed emission can overwhelm the atomic line radiation flux and might preferentially arise from high luminosity, high redshift BL Lac objects \cite{Gio11}. These would then be absent in the spectral index/luminosity diagram (Fig.\ \ref{fig:index_lum}) and skew the correlation.  Until the redshift incompleteness, the nature of the unassociated sources in the 2LAC, and underlying biases introduced by using different source catalogs are resolved, conclusions about the blazar sequence \cite{Fossati98,Ghi98} and the blazar divide \cite{Ghisellini09} remain tentative. It is also worthwhile to mention that the correlation visible for blazars in Figure~\ref{fig:index_lum} is very weak if FSRQs and BL~Lacs are considered separately.

\subsection{GeV-TeV connection}

\begin{figure}
\centering
\includegraphics[width=9cm]{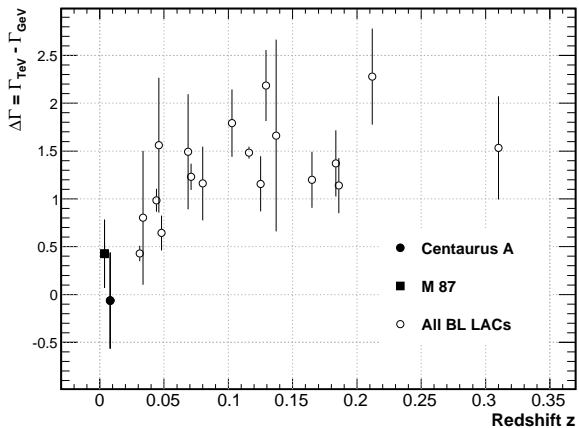}
\caption{The difference in photon index between that measured by {\it{Fermi}}
LAT and that reported in the TeV regime,
$\Delta\Gamma\,\equiv\,\Gamma_{TeV}\,-\,\Gamma_{GeV}$, for the 19
GeV-TeV AGN with reliable redshifts and reported TeV spectra as a function of their redshift.}
\label{fig:GeVTeVEBL}
\end{figure}

Out of 45 AGNs detected in the TeV regime and listed in the TeVCat catalog\footnote{http://tevcat.uchicago.edu}, 39 are in 2LAC and 34 are in the Clean Sample. {\it{Fermi}}-LAT was implicated in the detection of nine of these objects. Please note that only 13 TeV AGNs display significant variability in the GeV range. For the 2LAC Clean Sample, the largest subclass in the TeV 
AGNs (18) is the HSPs but there also 6 ISPs, 5 LSPs and 5 AGNs whose
SED class remains unclassified using the technique described above. The mean photon index of the 2LAC sources
associated with the TeV AGNs is $1.87\,\pm\,0.27$, while the mean
photon index of the Clean 2LAC Sample is $2.13\,\pm\,0.30$, indicating
that those AGNs which are detected at TeV are, in general, harder than
the majority of the 2LAC sources at {\it{Fermi}}-LAT energies.
For eight TeV AGNs in the Clean Sample; the {\it{Fermi}} LAT spectra are best fit by a LogParabola. In many cases, there is a significant
difference between the Power-law spectral indices measured by {\it{Fermi}}
LAT, $\Gamma_{GeV}$, and by the Cherenkov telescopes, $\Gamma_{TeV}$,
indicating that the spectrum undergoes a break somewhere in the
$\gamma$-ray regime (Fig. \ref{fig:GeVTeVEBL}). However, most of the TeV spectra have not been measured
simultaneously with the GeV spectra, and caution is advised when comparing the
spectra in detail.  The mean break index is $\langle\Delta\Gamma\rangle$=1.3. An apparent deficit of sources with small
$\Delta\Gamma$ is observed at high redshift which may partly be the effects of pair production with the EBL,
which is expected to introduce a redshift-dependent steepening into the TeV
spectra of extragalactic objects.

\section{Conclusion}

In the long-run, the Fermi-LAT will remain extremely valuable for monitoring variable sources or detecting transient events, as the sensitivity roughly scales as the square root of time while the number of detected transients scales linearly with time. 

\section{Acknowledgements}

The \textit{Fermi} LAT Collaboration acknowledges generous ongoing support from a number of agencies and institutes that have supported both the development and the operation of the LAT as well as scientific data analysis. These include the National Aeronautics and Space Administration and the Department of Energy in the United States, the Commissariat \`a l'Energie Atomique and the Centre National de la Recherche Scientifique / Institut National de Physique Nucl\'eaire et de Physique des Particules in France, the Agenzia Spaziale Italiana and the Istituto Nazionale di Fisica Nucleare in Italy, the Ministry of Education, Culture, Sports, Science and Technology (MEXT), High Energy Accelerator Research Organization (KEK) and Japan Aerospace Exploration Agency (JAXA) in Japan, and the K.~A.~Wallenberg Foundation, the Swedish Research Council and the Swedish National Space Board in Sweden.

\end{document}